\documentclass[12pt]{article}

\usepackage[english]{babel}
\usepackage{amsmath}
\usepackage{amssymb}
\usepackage{amsfonts}

\begin{document}

\newcommand{\be}{\begin{equation}}
\newcommand{\ee}{\end{equation}}
\newcommand{\bea}{\begin{eqnarray}}
\newcommand{\eea}{\end{eqnarray}}
\newcommand{\beaa}{\begin{eqnarray*}}
\newcommand{\eeaa}{\end{eqnarray*}}
\newcommand{\Lhat}{\widehat{\mathcal{L}}}
\newcommand{\nn}{\nonumber \\}
\newcommand{\e}{\mathrm{e}}
\newcommand{\tr}{\mathrm{tr}\,}

\begin{center}

{\Large\bf Viscous Little Rip Cosmology}
\vspace{5mm}

I.~Brevik$^{a,}$\footnote{E-mail: iver.h.brevik@ntnu.no},
E.~Elizalde$^{b,}$\footnote{E-mail: elizalde@ieec.uab.es and elizalde@math.mit.edu},
S.~Nojiri$^{c,d,}$\footnote{E-mail: nojiri@phys.nagoya-u.ac.jp}
 and S.~D. Odintsov$^{b,e,f,}$\footnote{E-mail: odintsov@ieec.uab.es}
\vspace{3mm}

{\small
$^a$Department of Energy and Process Engineering,
\\
Norwegian University of Science and Technology, N-7491 Trondheim, Norway
\\
$^b$Consejo Superior de Investigaciones Cient\'{\i}ficas, ICE/CSIC-IEEC,
\\
Campus UAB, Facultat de Ci\`{e}ncies, Torre C5-Parell-2a pl, E-08193
Bellaterra (Barcelona) Spain
\\
$^c$Department of Physics, Nagoya University, Nagoya 464-8602, Japan
\\
$^d$Kobayashi-Maskawa Institute for the Origin of Particles and the Universe,
\\
Nagoya University, Nagoya 464-8602, Japan
\\
$^e$Instituci\'{o} Catalana de Recerca i Estudis Avan\c{c}ats (ICREA)
\\
$^f$Tomsk State Pedagogical University

\bigskip
Revised version, October 13, 2011
}

\vspace{3mm}

\begin{abstract}

Dark energy of phantom or quintessence nature with an equation of state parameter $w$
almost equal to $-1$  often leads the universe evolution to a finite-time future
singularity. An elegant solution to this problem has been recently proposed \cite{frampton11} under the form of the so-called Little Rip cosmology which appears
to be a realistic alternative to the $\Lambda$CDM model. A viscous Little Rip cosmology is here proposed. Whereas generically bulk viscosity tends to promote the Big Rip, we find that there are a number of situations where this is not the case and where the formalism nicely adjusts itself to the Little Rip scenario. We prove, in particular, that a viscous fluid (or, equivalently, one with an inhomogeneous (imperfect) equation of state) is perfectly able to produce a Little Rip cosmology as a purely viscosity effect. The possibility of its induction as a combined result of viscosity and a general (power-like) equation of state is also investigated in detail. To finish, a physical, inertial force interpretation of the dissolution of bound structures in the Little Rip cosmology is presented.

\end{abstract}

\end{center}

\noindent {\it 1. Introduction.}---Viscous cosmology models have been increasingly popular lately. From a hydrodynamical viewpoint this is quite a natural development, as the inclusion of the  viscosity coefficients (there are two of them, shear and bulk) means physically that one departs from the case of an ideal fluid and incorporates the deviations from thermal equilibrium to first order. The case of an ideal (nonviscous) fluid is, after all, quite an idealized model, useful in practice for many situations but not so for all, especially not when fluid motion near solid boundaries is involved. Also under boundary-free conditions--such as in free turbulence---viscosities are physically most important. In a cosmological context, as the cosmic fluid is assumed to be spatially isotropic the shear viscosity is usually left out. Anisotropic deviations such as those encountered in the Kasner universe are quickly smoothed out being the only coefficient left $\zeta$, the bulk viscosity.
Some of the early treatises on viscous cosmology are \cite{weinberg71}--\cite{brevik94}. Recently, work has aimed at the study of the dark energy universe when the dark fluid is a viscous one \cite{cataldo05}. In \cite{nojiri05} a dark fluid with a time dependent bulk viscosity was considered as a fluid obeying an inhomogeneous (imperfect) equation of state.

One place where the appearance of bulk viscosity in the cosmic fluid should be expected to play an important role, is in the Big Rip (BR) phenomenon, i.e. the singularity of the universe in the future (for pioneering papers on the BR, see \cite{caldwell03,mcinnes02}). The Big Rip (or type II, III and IV Rip \cite{barrow04,nojiri05A}) means that  one or more of the physical quantities go to infinity at a finite time $t$ in the future. Mathematically this implies divergent integrals, typically met when one uses the Friedmann equations to express  $t$ as an integral over the density $\rho$. For phantom/quintessence dark energy models which develop a finite-time future singularity of one of the four known types
\cite{nojiri05A}, the following scenarios to avoid future singularities were proposed: quantum effects \cite{Elizalde:2004mq}, modified gravity \cite{bno1}, and coupling with dark matter \cite{noplb1}.

Recently, a novel scenario has been proposed in \cite{frampton11}, the so-called ``Little Rip" (LR). Models are there examined in which the nonviscous dark energy  density increases with time (equation of state parameter $w<-1$), but $w\rightarrow -1$ asymptotically, in that way avoiding the future singularity. Typically, this was found to occur when the scale factor increases rapidly with time, like $a(t) \sim \exp(\exp(t))$ or higher exponentials. And this brings us to the topic of the present paper, namely to examine the consequences of endowing the fluid with a bulk viscosity. In most cases, the presence of a bulk viscosity tends to promote the Big Rip phenomenon; an example being presented at the end of section 2. However, it turns out that the formalism can nicely fit the Little Rip scenario under special conditions. One such example will be considered in detail in section 4. Particular underlying conditions for this case to occur will be thoroughly discussed for a large class of models. Specific examples are when the bulk viscosity is constant, or is inversely proportional to the Hubble parameter. A mixed effect of viscosity and a generalized (power-like) equation of state in the non-viscous sector can also lead to a LR cosmology. It will be explicitly demonstrated that, for the standard equation of state $p=w\rho$, just taking into account the special viscous sector may bring the universe evolution to a LR. To finish, an inertial force interpretation of the dissolution of bound structures in the LR cosmology will be developed.

 We may here notice  that there is in general a problem with a possible transient acceleration in the universe: the acceleration may turn back to a usual deceleration regime in the future. See, for instance, the  papers \cite{albrecht00} discussing this theme. In principle, one may expect such scenarios also in modified gravity. In our present model, however, the Little Rip cosmology turns out to be stable (effectively the phantom-like regime), so we do not expect such a transient acceleration behavior in the model under consideration.

\medskip

\noindent {\it 2. Basic formalism: General viscosity.}---We assume first that that the cosmic fluid possesses a shear viscosity $\eta$ in addition to the bulk viscosity $\zeta$. We put $k_B=c=1$, take the Minkowski metric in the form (-+++), sum Greek indices from 0 to 3, and Latin indices from 1 to 3. If $U^\mu=(U^0,U^i)$ denotes the four-viscosity of the fluid, we have in comoving coordinates $U^0=1, U^i=0$.

Let $g_{\mu\nu}$ be a general metric, and introduce the projection tensor
$
h_{\mu\nu}=g_{\mu\nu}+U_\mu U_\nu. 
$
Then the rotation and expansion tensors can be expressed, respectively, as
\begin{equation}
\omega_{\mu\nu}=\frac{1}{2}(U_{\mu;\alpha}h_\nu^\alpha-U_{\nu;\alpha}h_\mu^\alpha),  \qquad 
\theta_{\mu\nu}=\frac{1}{2}(U_{\mu;\alpha}h_\nu^\alpha+U_{\nu;\alpha}h_\mu^\alpha), \label{3}
\end{equation}
being the scalar expansion $\theta=\theta_\mu^\mu={U^\mu}_{;\mu}$. In a FRW space $\theta=3H$, with $H$ the Hubble parameter. This relationship will be made use of in the following. The third tensor to be defined is the shear tensor,
\begin{equation}
\sigma_{\mu\nu}=\theta_{\mu\nu}-Hh_{\mu\nu}. \label{4}
\end{equation}
It is traceless, $\sigma_\mu^\mu=0$. The following decomposition of the covariant derivative $U_{\mu;\nu}$ is often useful,
\begin{equation}
U_{\mu;\nu}=\omega_{\mu\nu}+\sigma_{\mu\nu}+Hh_{\mu\nu}-A_\mu U_\nu, \label{5}
\end{equation}
where $A_\mu=\dot{U}_\mu=U^\alpha U_{\mu;\alpha}$ is the four-acceleration of the fluid.

Consider now the fluid's energy-momentum tensor $T_{\mu\nu}$, assuming for simplicity that the temperature $T$ is constant,
\begin{equation}
T_{\mu\nu}=\rho U_\mu U_\nu+  (p-3\zeta H)  h_{\mu\nu}-2\eta \sigma_{\mu\nu}. \label{6}
\end{equation}
Here
\begin{equation}
p_{\rm eff}=p-3\zeta H \label{7}
\end{equation}
 is the effective pressure, lower than $p$ because of the inequality $\zeta \geq 0$ which in turn is a consequence of thermodynamics.

 The entropy per unit comoving volume is $S=n\sigma$, where $n$ is the particle number density and $\sigma$ the entropy per particle. This corresponds to an entropy current four-vector and corresponding four-divergence
 \begin{equation}
 S^\mu=n\sigma U^\mu, \label{8} \qquad
 {S^\mu}_{;\mu}=\frac{2\eta}{T}\sigma_{\mu\nu}\sigma^{\mu\nu}+\frac{9\zeta}{T}H^2. 
 \end{equation}
 Thus, both viscosity coefficients, $\eta$ and $\zeta$, contribute to the heat production in the fluid.

 We now specialize to the case of an isotropic fluid, $\eta=0$, and introduce the equation of state
 \begin{equation}
 p=-\rho-f(\rho), \label{10}
 \end{equation}
 $f(\rho)$ being an arbitrary function. What is most important in viscous cosmology, is not $p$ but instead the effective pressure $p_{\rm eff}$. We generalize the expression (\ref{7}), assuming that $\zeta(H)$ is a general function of $H$. Then, with $3\zeta H \rightarrow \xi(H)$ we express the effective pressure as
 \begin{equation}
 p_{\rm eff}=-\rho-f(\rho)-\xi(H), \label{11}
 \end{equation}
which is also called cosmological inhomogeneous (imperfect) equation of state \cite{nojiri05} (in general one should allow the function $\xi$ to depend on derivatives of $H$ too).

Consider now the spatially flat FRW line element
 \begin{equation}
 ds^2=-dt^2+a^2(t)d{\bf x}^2.
 \end{equation}
 The first and second Friedmann equations are, respectively ($H=\dot{a}/a$),
 \begin{equation}
 H^2=\frac{8\pi}{3}G\rho,  \qquad  
 \frac{\ddot{a}}{a}+\frac{1}{2}H^2=4\pi G [\rho+f(\rho)+\xi(H)]  \label{14}
 \end{equation}
 (note that it is essentially the effective pressure that appears on the right hand side).
 From the conservation of energy, ${T^{0\nu}}_{;\nu}=0$, we get
 \begin{equation}
 \dot{\rho}-3f(\rho)H=    3\xi(H)H. \label{15}
 \end{equation}
 The special case where $\zeta(\rho)$ is set proportional to $H$ is of interest:
  \begin{equation}
 \zeta(\rho)=3\tau H=\tau\sqrt{24\pi G\rho}, \label{16}
 \end{equation}
 with $\tau$  the proportionality constant. The ansatz (\ref{16}) should be quite reasonable physically, as it implies that the viscosity, i.e. the deviation from thermal equilibrium, is most pronounced when the motion of the universe is most violent. In this case, we obtain
 \begin{equation}
t=\frac{1}{\sqrt{24\pi G}}\int_{\rho_0}^\rho \frac{d\rho}{\rho^{3/2}\left[f(\rho)/\rho+24\pi G\tau \right]}. \label{17}
\end{equation}
 Here the integration starts from $t_0=0$, the present time. The Big Rip is inevitable if this integral is finite for $\rho \rightarrow \infty$.

 It is instructive to make the ansatz \cite{nojiri05,frampton11}
 \begin{equation}
 f(\rho)=A\rho^\alpha, \label{18}
 \end{equation}
 where $A$ and $\alpha$ are constant. If the fluid is nonviscous, $\tau=0$, it is seen that the value $\alpha=1/2$ represents a borderline between Big Rip and non-Big Rip. If $\alpha < 1/2$, the singularity is avoided. However, for a viscous fluid the situation is different: even if $f(\rho)/\rho \rightarrow 0$ for $\rho \rightarrow \infty$ the integral reduces effectively to $\int \rho^{-3/2}d\rho \propto \rho^{-1/2}$ for large $\rho$.

 We thus arrive at the following important conclusion: if $\zeta \propto H$ the Big Rip is inevitable, regardless of the form of the function $f(\rho)$.
\medskip

\noindent {\it 3. Little Rip cosmology: no viscosity.}---We first reproduce some essentials from Ref.~\cite{frampton11}. When $\xi(H)=0$ it follows that the equation $H=\dot{\rho}/(3f(\rho))$ is satisfied by the following integral expression for the scale factor:
 \begin{equation}
 a=\exp \left(\int_{\rho_0}^\rho \frac{d\rho}{3f(\rho)} \right). \label{19}
 \end{equation}
 We have here put the scale factor at present time, $a_0$, equal to unity. The first Friedmann equation then yields
 \begin{equation}
 t=\frac{1}{\sqrt{24\pi G}}\int_{\rho_0}^\rho \frac{d\rho}{\sqrt{\rho}f(\rho)}. \label{20}
 \end{equation}
 We assume hereafter that $f(\rho)$ is given by the power law (\ref{18}), and put $\alpha=1/2$. Then $f(\rho)=A\sqrt{\rho}$, and
 \begin{equation}
 p=-\rho-A\sqrt{\rho}. \label{21}
 \end{equation}
 From Eq.~(\ref{17}) we obtain in the nonviscous case
 \begin{equation}
 t=\frac{1}{\sqrt{24\pi G}}\,\frac{1}{A}\,\ln \frac{\rho}{\rho_0}, \label{22}
 \end{equation}
 showing that $\rho \rightarrow \infty$ is not reached until an infinite time $t$ has elapsed. This is the Little Rip phenomenon. Note that it is called this way because it also produces disintegration of bound structures, as we discuss latter.

 Using Eq.~(\ref{19}) we can also express $\rho$ as a function of $a$,
 \begin{equation}
 \rho(a)=\rho_0 \left( 1+\frac{3A}{2\sqrt{\rho_0}}\ln a \right)^2. \label{23}
 \end{equation}
 Again using the first Friedmann equation we get
 \begin{equation}
 \int_1^a \frac{da/a}{1+\frac{3A}{2\sqrt{\rho_0}}\ln a}=\sqrt{\frac{3}{8\pi G\rho_0}}\,t, \label{24}
 \end{equation}
 which upon integration permits us to find the scale factor as a function of time,
 \begin{equation}
 a(t)=\exp \left\{ \frac{2\sqrt{\rho_0}}{3A}\left[\exp \left(\frac{3A}{2\rho_0}\sqrt{\frac{3}{8\pi G}}\,t\right)-1 \right] \right\}. \label{25}
 \end{equation}
 As explained in Ref.~\cite{frampton11}, this LR cosmology is effectively non-singular, because the singularity only occurs in the infinite future.
\medskip

\noindent {\it 4. Little Rip cosmology: constant viscosity function $\xi(H)$.}---Now, we move on to consider the viscosity-dependent governing equations. Friedmann's first equation in (\ref{14}) remains formally unchanged, while the second and (\ref{15}) become altered. It is seen that the formalism becomes easily manageable if the following condition on the viscosity holds,
 \begin{equation}
 \xi(H)=\xi_0= \rm const. \label{26}
 \end{equation}
 In that case we obtain from Eq.~(\ref{15})
 \begin{equation}
 a=\exp \left( \int_{\rho_0}^\rho \frac{d\rho}{3[\xi_0+f(\rho)]}\right), \label{27}
 \end{equation}
 and further use of the first Friedmann equation then yields
 \begin{equation}
 t=\frac{1}{\sqrt{24\pi G}}\int_{\rho_0}^\rho \frac{d\rho}{\sqrt{\rho}\,[\xi_0+f(\rho)]}. \label{28}
 \end{equation}
 If we assume $f(\rho)=A\sqrt{\rho}$ as before (it implies that the dimension of $A$ in geometric units is cm$^{-2}$), we get
 \begin{equation}
 t=\frac{1}{\sqrt{6\pi G}} \frac{1}{A}\ln \frac{\xi_0+A\sqrt{\rho}}{\xi_0+A\sqrt{\rho_0}}. \label{29}
 \end{equation}
 This equation can be inverted to give $\rho$ as a function of $t$,
 \begin{equation}
 \rho(t)=\left[ \left(\frac{\xi_0}{A}+\sqrt{\rho_0}\right) \exp (\sqrt{6\pi G} At)-\frac{\xi_0}{A}\right]^2. \label{30}
 \end{equation}
 Thus $\rho \rightarrow \infty$ can be reached, but the universe needs an infinite time to do so. This is precisely the Little Rip phenomenon, now met under  viscous conditions. Recall again that this conclusion rests upon the equation of state being adopted in the form (\ref{21}). We see that $\rho (t)$ increases more rapidly in the presence of viscosity. It may be interesting to compare it with the turbulence effect to dark energy
 \cite{bgno10}.

 The effective parameter determining the influence from viscosity is the nondimensional quantity $\varepsilon=\xi_0/(A\sqrt{\rho_0})$. A weak influence from viscosity, which is probably the case most expected, implies that $\varepsilon \ll 1$. The nondimensional quantity $\xi_0/(A\sqrt{\rho(t)})$ will rapidly become much less than $\varepsilon$ as  time increases.
\medskip

\noindent {\it 5. More general cases.}---The above analysis can be extended to much more general cases, concerning the dependencies of both
the function $f(\rho)$ and the bulk viscosity, $\zeta$, in terms of the density $\rho$ (or Hubble constant $H$, see Eq.~(\ref{16})). The most general situation which will be here discussed is:
 \begin{equation}
 f(\rho)=A \, \rho^{\beta+1/2}, \qquad  \zeta = b \, \rho^\gamma.  \label{31a}
 \end{equation}
 Note that the particular case studied in the previous section corresponds, in this notation, to $\gamma=-1/2$, $\beta=0$. Proceeding as before, we now obtain for Eq.~(\ref{28})
 \begin{equation}
 t=\frac{1}{\sqrt{24\pi G}}\int_{\rho_0}^\rho \frac{d\rho}{\rho \,(A \, \rho^\beta + b \, \rho^\gamma)}. \label{32a}
 \end{equation}
This expression can be integrated generically in terms of an hypergeometric function, the result being
 \begin{eqnarray}
 t&=&\frac{1}{\sqrt{24\pi G}} \left\{ \frac{\beta \, \rho^{-\gamma}}{\gamma b (\gamma-\beta)} + \left.
 \frac{\rho^{-\gamma}}{\gamma b (\gamma-\beta)} \right[ -\beta  \right. \nonumber \\ && \left. \left. + (\beta -\gamma)\, {}_2F_1 \left( 1, \frac{\gamma}{\gamma-\beta}, 1 + \frac{\gamma}{\gamma-\beta}, -\frac{A}{\beta}\rho^{\beta -\gamma}\right)\right]\right\}. \label{33a}
 \end{eqnarray}

Some particular cases deserve special attention. When the bulk viscosity is constant (that is, $\gamma=0$) and also $\beta =0$ (as in the previous section), we obtain again the result that $t$ has a logarithmic dependence on $\rho$
\begin{equation}
 t=\frac{1}{\sqrt{24\pi G} \, (A+b)}\ln \frac{\rho}{\rho_0}, \label{34a}
 \end{equation}
and, proceeding as above, the Little Rip phenomenon shows up. Notice that this was obtained before for the case when the bulk viscosity was inversely proportional to the Hubble constant (or, what is the same, to the square root of the density), while here it is the bulk viscosity itself which is strictly constant. In fact, it is easy to see that this result is more general: keeping  $\beta =0$, for any $\gamma \leq 0$ we have a LR, while for $\gamma > 0$ a BR occurs.
It is plain that a similar situation appears in the opposite case: when $\gamma=0$, for $\beta \leq 0$ we have a LR, while for $\beta > 0$ a BR.

In the subclass of cases when $\gamma=\beta > 0$ one obtains that the value $\rho = \infty$ is reached at a finite and positive time, namely,
\begin{equation}
 t_{BR}=\frac{\rho_0^{-\gamma}}{\sqrt{24\pi G} \, (A+b)\, \gamma}, \label{35a}
 \end{equation}
so that we are back to the Big Rip situation in all these cases. Things change, however, when
$\gamma=\beta < 0$, in which circumstance no singularity can be ever reached in finite time.

An asymptotic analysis of the general formula (\ref{33a}), which encompasses the whole class of dependencies here discussed (\ref{31a}), yields a explicit result for each particular situation one may be interested in, within the three possibilities: absence of a Rip singularity (e.g., $\gamma >\beta$, with $\gamma > 0$), eventual formation of a Little Rip singularity (that is, one that is reached asymptotically at infinite time), and formation of a Big Rip singularity at a finite time in the future (e.g., $\gamma >\beta$, with $\gamma < 0$). The interplay between the dependencies of both the function $f(\rho)$ and the bulk viscosity, $\zeta$, in terms of the density $\rho$ (or Hubble constant $H$), in getting a specific cosmology in which one or another of these possibilities is realized, is explicitly manifest in the calculation carried out in this section.
\medskip

\noindent {\it 6. On the physics of the viscous LR.}---In order to better understand the influence of cosmological viscosity in physical terms, we start again from Eq.~(\ref{11}) but now, for simplicity,
\be
\label{h2}
p = w \rho - \xi(H)\, ,
\ee
that is, the ordinary EoS just modified by addition of the viscosity term. The conservation of the energy density reads
\be
\label{h3}
0 = \dot \rho + 3 H \left[ \left( 1 + w \right)\rho - \xi(H) \right]\, .
\ee
Consider now the following Little Rip model:
\be
\label{h4}
H= H_0 \e^{\lambda t}\, ,
\ee
with $H_0$ and $\lambda$  positive constants.
The parameter $A$ in \cite{frampton11} and Eq.~(\ref{18}) corresponds to $\sqrt{2}\lambda$ in (\ref{h4}) and we know it is bounded as $2.74\times 10^{-3}\, \mathrm{Gyr}^{-1} \leq A \leq 9.67\times 10^{-3}\, \mathrm{Gyr}^{-1}$ by the Supernova Cosmology Project observational data \cite{Amanullah:2010vv}.
In \cite{frampton11}, the model (\ref{h4}) gives almost identical behavior of the distance versus redshift and, therefore, the simple model (\ref{h4}) turns out to be very realistic.

By using the first FRW equation, we find
\be
\label{h5}
\rho = \frac{3}{\kappa^2} H^2 = \frac{3}{\kappa^2} H_0^2 \e^{2\lambda t}\, ,
\ee
and therefore
\be
\label{h6}
\dot\rho = \frac{6\lambda}{\kappa^2} H_0^2 \e^{2\lambda t} = \frac{6\lambda}{\kappa^2} H^2 \, .
\ee
Then, from Eq.~(\ref{h3})
\be
\label{h7}
0 = \frac{6\lambda}{\kappa^2} H^2 + 3H \left[ \frac{3\left(1+w\right)}{\kappa^2}H^2
 - \xi(H) \right] \, ,
\ee
which can be solved with respect to $\xi(H)$ as
\be
\label{h8}
\xi(H) = \frac{3\left(1+w\right)}{\kappa^2}H^2 + \frac{2\lambda}{\kappa^2} H\, .
\ee
Conversely, if we start with a perfect fluid whose equation of state is given by
(\ref{h2}) and (\ref{h8}), there is a solution which realizes the LR as (\ref{h4}).
The remarkable point is that such LR is induced purely by the viscosity sector.

Instead of (\ref{h4}), we may now consider
\be
\label{g1}
H = H_0 \e^{C \e^{\lambda t}}\, .
\ee
Here $H_0$, $C$ and $\lambda$ are positive constants.
It has been shown in \cite{frampton11} that this model (\ref{g1}) can be realistic too.
We find $\rho$ and $\dot\rho$ as
\be
\label{h9}
\rho = \frac{3}{\kappa^2} H_0^2 \e^{2C \e^{\lambda t}}\, , \qquad
\dot \rho = \frac{6C\lambda}{\kappa^2} H_0^2 \e^{2C \e^{\lambda t}} \e^{\lambda t}
= \frac{6\lambda}{\kappa^2} H^2 \ln \frac{H}{H_0}\, .
\ee
while $\xi(H)$ has the following form:
\be
\label{h10}
\xi(H) = \frac{3\left(1+w\right)}{\kappa^2}H^2 + \frac{2\lambda}{\kappa^2} H
\ln \frac{H}{H_0}\, .
\ee
Using the first FRW equation, $\rho = \frac{3}{\kappa^2} H^2 $, the conservation law
(\ref{h3}) can be rewritten as
\be
\label{h11}
0 = \frac{2}{\kappa^2}\dot H + 3 \left[ \frac{3\left(1+w\right)}{\kappa^2} H^2 - \xi (H) \right] \, .
\ee
In the case of (\ref{h8}), we obtain
\be
\label{h12}
0=\dot H - \lambda H\, ,
\ee
whose unique solution is (\ref{h4}) up to the constant shift of the time coordinate $t\to t + t_0$.
On the other hand, in the case of (\ref{h10}), we get
\be
\label{h13}
0=\dot H - \lambda H \ln \frac{H}{H_0} \, .
\ee
For this Eq.~(\ref{h13}), besides the solution (\ref{g1}),  the de Sitter solution $H=H_0$ appears,
which can be regarded as a special case corresponding to $C\to 0$ in (\ref{g1}).
The stability of the de Sitter solution can be now investigated by performing the perturbation
\be
\label{h14}
H = H_0 + \delta H \, .
\ee
Then, Eq.~(\ref{h13}) yields
\be
\label{h15}
0 = \delta \dot H - \lambda \delta H\, .
\ee
We find that $\delta H$ grows exponentially $\delta H \propto \e^{\lambda t}$ and,
therefore, the de Sitter solution $H=H_0$ is unstable. This indicates  that the universe could actually evolve into the LR solution (\ref{g1}).

When the universe is expanding, the relative acceleration between two points, whose distance is $l$, is given by
$l \ddot a/a$ ($a$ is the scale factor). A particle with mass $m$ at a certain point will be subject to an inertial force
\be
\label{i1}
F_\mathrm{inert}=m l \ddot a/a = m l \left( \dot H + H^2 \right)\, ,
\ee
whenever it is observed from another point.
Let us assume the two point force is bounded by the bounding force $F_0$. If $F_\mathrm{inert}$ is positive and bigger than $F_0$, the bound on the two points is destroyed. This is a ``Rip'' caused by the accelerating expansion.
The ``Rip'' always occurs when $H$ diverges, which corresponds to a ``Big Rip'' \cite{caldwell03}.
Even if $H$ is finite, if $\dot H>0$
diverges, which corresponds to a Type II or ``sudden future'' singularity
\cite{barrow04,nojiri05A},  the strong rip will always occur.
Even if $H$ is finite in finite future, in the case when $H$ becomes infinite in the infinite future,
the inertial force (\ref{i1}) becomes larger and larger and destroy any bounded object. This has been dubbed as ``Little Rip'' \cite{frampton11}.
In this case and for realistic models, the time $t_\mathrm{LR}$ from present until the destruction of the  Earth-Sun system  has been estimated.
In the case of (\ref{h4}) it amounts to $t_\mathrm{LR} \sim 8\,$Tyrs and in the case of (\ref{g1}) to $t_\mathrm{LR} \sim 146\,$Gyrs.

In summary, we have here considered the role of a viscous (or inhomogeneous (imperfect)
equation of state) fluid in a Little Rip cosmology.
Despite the earlier observations that viscosity basically supports the Big Rip
singularity, we have demonstrated that it is also able to give rise to a non-singular, Little Rip cosmology, which is considered to be a viable alternative to $\Lambda$CDM
cosmology. In particular, constant bulk viscosity and a viscosity inversely proportional to the Hubble rate have been considered. We have shown that in those cases a Little Rip cosmology may naturally emerge. It is remarkable that for a standard fluid, $p=w\rho$, the only influence of viscous effects can naturally drive the universe to a Little Rip type evolution. What is more, in some cases the Little Rip cosmology may be even more stable than de Sitter space itself. This invites to a very careful investigation of all properties which are specific of the Little Rip cosmology, which avoids the well-known problems of the Big Rip universe.
\medskip

\noindent {\it Acknowledgments.}---We are grateful to P. Frampton and R. Scherrer for stimulating discussions. EE and SDO are grateful to K. Olaussen for kind hospitality at NTNU where this work was completed. This research has been supported by the ESF CASIMIR Network, by MICINN (Spain) projects FIS2006-02842 and FIS2010-15640,
by CPAN Consolider Ingenio Project and AGAUR 2009SGR-994 (EE and SDO),
and by the Global COE Program of Nagoya University (G07) provided by the Ministry of
Education, Culture, Sports, Science \& Technology of Japan
and the JSPS Grants-in-Aid for Scientific Research (S) \# 22224003 and
(C) \#23540296 (SN).

\end{document}